\documentclass[a4paper,showkeys]{emulateapj}

\usepackage{amsmath,amssymb,amsfonts}
\usepackage{color}

\usepackage[T1]{fontenc}
\usepackage{txfonts}
\usepackage[]{graphicx}

\shorttitle{Fake massive black holes in the LISA band}
\shortauthors{Xian Chen, Ze-Yuan Xuan, \& Peng Peng, 2019}

\begin{document}

\title{Fake massive black holes in the milli-Hertz gravitational-wave band}

\author{Xian Chen}
\email{Corresponding author: xian.chen@pku.edu.cn}
\affiliation{Astronomy Department, School of Physics, Peking University, 100871 Beijing, China}
\affiliation{Kavli Institute for Astronomy and Astrophysics at Peking University, 100871 Beijing, China}

\author{Ze-Yuan Xuan}
\affiliation{Physics Department, School of Physics, Peking University, 100871 Beijing, China}

\author{Peng Peng}
\affiliation{Astronomy Department, School of Physics, Peking University, 100871 Beijing, China}

\date{\today}

\begin{abstract}
In gravitational wave (GW) astronomy accurate measurement of the source
parameters, such as mass, relies on accurate waveform templates.  Currently,
the templates are developed assuming that the source, such as a
binary black hole (BBH), is residing in a vacuum. However, astrophysical models
predict that BBHs could form in gaseous environments, such as common envelops,
stellar cores, and accretion disks of active galactic nuclei.  Here we revisit
the impact of gas on the GW waveforms of stellar-mass BBHs  with a focus on the
early inspiral phase when the GW frequency is around milli-Hertz.  We show that
for these BBHs, gas friction could dominate
the dynamical evolution and hence duplicate chirp signals. The
relevant hydrodynamical timescale, $\tau_{\rm gas}$, could be much shorter than the GW radiation timescale,
$\tau_{\rm gw}$, in the above astrophysical scenarios. As a result, the observable chirp  mass 
	is higher than the real one by a
factor of $(1+\tau_{\rm gw}/\tau_{\rm gas})^{3/5}$ if the gas effect is
ignored in the data analysis. Such an error also results in an overestimation
of the source distance  by a factor of $(1+\tau_{\rm gw}/\tau_{\rm gas})$. 
By performing matched-filtering analysis in the milli-Hertz band,
we prove that the gas-dominated signals are practically
indistinguishable from the chirp signals of those more massive BBHs residing in 
a vacuum environment.  Such fake massive objects in the milli-Hertz band, if not
appropriately accounted for in the future, may alter our understanding of the formation,
evolution, and detection of BBHs.
\end{abstract}


\section{Introduction}\label{sec:intro}

The majority of the black holes (BHs) detected by the ground-based
gravitational wave (GW) observatories (i.e, LIGO and Virgo) turn out to be
several times more massive than those previously detected in X-ray binaries
\citep{ligo18a}.  Such a discrepancy has important implications for the
formation and evolution of stellar-mass binary BHs
\citep[BBHs,][]{ligo16astro}. However, mass is not a direct observable in GW
astronomy.  It is inferred either from the chirp signal, i.e., an increase of
the GW frequency with time, or from the merger and ringdown signals.  If the
signal gets distorted, either at the moment of generation or during the
propagation, an error would be induced in the measurement of the mass.

Redshift is such a disturbing factor. It stretches the signal during its
propagation.  As a result, mass ($m$) is degenerate with redshift ($z$) so that
one can only measure from GW signals the redshifted mass $m(1+z)$
\citep{schutz86}.  In two astrophysical scenarios, the mass-redshift degeneracy
could lead to an significant overestimation of the masses of BBHs. In the first
scenario, a BBH is at a high cosmological redshift and, nevertheless, is
detected because the GW signal is magnified due to gravitational lensing
\citep{broadhurst18,smith18}.  So far, there is no convincing evidence
supporting this scenario \citep{hannuksela19}.  This scenario also has
difficulties explaining the positive correlation between the apparent masses
and distances of the detected BBHs \citep{ligo18a}. In the second scenario, the
BBH is captured by a supermassive black hole (SMBH) to a small distance, so
that Doppler and gravitational redshifts become significant, which then lead to
the mass-redshift degeneracy \citep{chen_han_2018,chen19}. The uncertainty of
this latter scenario lies mainly in the poor understanding of the event rate.
In both scenarios the signal in the LIGO/Virgo band (centered around $10-10^2$
Hz) is similar to that of a not-redshifted, but more massive BBH.
Distinguishing them would be difficult partly because the signal is short,
which normally lasts no more than one second, too short to reveal any signature
of gravitational lensing \citep{hannuksela19} or a nearby SMBH \citep{chen19}.

The difficulty would be alleviated if the BBH can be detected by a space-borne
GW observatory, such as the Laser Interferometer Space Antenna \citep[LISA,
][]{danzmann_2017}. LISA is sensitive to milli-Hertz (mHz) GWs and hence can
capture a BBH at the early inspiral phase, weeks to millennia before it enters
the LIGO/Virgo band \citep{miller02,sesana16,moore19}. Using the chirp signal,
LISA can also measure the (redshifted) mass of the BBH.  In this way, LISA may
detect hundreds of massive BBHs during its mission duration of $4-5$ years
\citep{sesana16,kyutoku16,lamberts18,kremer18} and compare their masses with
those from LIGO/Virgo observations. 

Moreover, by tracking the BBHs with a decent signal-to-noise ratio (SNR, e.g.,
$>10$ ) for several months to years, LISA may reveal multiple images of a GW
source if it is strongly lensed \citep{seto04,sereno11}. The long signal may
also reveal a shift of the GW phase caused by the wave effect of gravitational
lensing \citep{nakamura98,takahashi03lensing}.  Moreover, if a BBH is close to
a SMBH, the long waveform should also contain imprints of the orbital motion of
the binary around the SMBH
\citep{inayoshi17,meiron17,robson18,chamberlain19,tamanini19,wong19,torres-orjuela20}
or the perturbation of the binary orbit by the tidal force of the SMBH
\citep{meiron17,hoang19,randall19,fang19}.  These signatures can help us
identify the BBHs affected by the redshift effects.

Redshift is not the only factor in GW astronomy that could affect the
measurement of mass. Gas, for example, can exert a frictional force on a
binary and hence lead to a faster orbital decay \citep{ostriker99,kim07,kim08}.
The resulting GW signal is expected to deffer from the real chirp signal due to
GW radiation only.
The impact on the mass measurement deserves further investigation
since a large fraction of BBHs may form in gaseous environments.
For example, BBHs can be produced by binary-star evolution, and in this case
the mergers may happen inside a common envelope
\citep{ivanova13,macleod17,ginat19} or the fallback material from the previous
supernovae \citep{tagawa18}.  Moreover, some BBHs may form in the accretion
disks of active galactic nuclei \citep[AGNs,][]{mckernan12,bartos17,stone17}.
When these BBHs merge, it is likely that they are surrounded by dense gas.
Furthermore, the dense cores of massive stars may also produce BBHs, and hence
the mergers would also be accompanied by gas \citep{loeb16,fedrow17,dorazio18}.
The density of the gas can reach $10^{8}-10^{14}\,{\rm cm^{-3}}$ in the case of
AGN disks, $10^{16}-10^{19}\,{\rm cm^{-3}}$ in common envelops, and even higher
in stellar cores \citep[see][ for a summary]{antoni19}.

Several earlier works studied the impact of gas on the GW signal of merging
binaries, including BBHs in the LIGO/Virgo band
\citep{fedrow17,ginat19,cardoso19} and extreme-mass-ratio inspirals (another
type of binary composed of a stellar-mass BH orbiting a SMBH) in the LISA band
\citep[][]{yunes11,kocsis11,barausse14}.  They focused on the final
evolutionary stage when the semi-major axes ($a$) of the binaries are only
$10-10^2$ times the Schwarzschild radius ($r_S$) of the bigger BHs. At this
stage, GW radiation predominates and gas plays a minor role. Nevertheless,
these works showed that gas could induce a small phase shift to the GW signal.
The phase shift does not significantly affect the mass measurement but can be
used to identify the mergers happening in gas.

The BBHs in the LISA band are very different from those considered in the
earlier works.  They have much greater semi-major axes
\citep[$a\sim10^3-10^4\,r_S$, e.g.,][]{chen17} since GW frequency
is proportional to $a^{-3/2}$ .  Such wide binaries have much weaker GW radiation
because the GW power scales with $a^{-5}$ \citep{peters64}. Consequently, gas
could play a dominant role in the evolution of these binaries and, in this way,
produce a fake chirp signal. If this factor is not accounted for in the LISA
data analysis, one may overestimate the masses, as we have shown in one example
of our preliminary study \citep[][Paper I]{chen19gas}.

Here we present our full analysis of the problem and discuss the detectability
of such ``fake'' massive BBHs in the LISA band. The paper is organized as
follows. In \S\ref{sec:theory} we explain how mass is measured from a chirp
signal and why it is affected by gas friction. Then in \S\ref{sec:compare} we
show that in realistic astrophysical scenarios gas friction indeed can overcome
GW radiation and dominate the orbital evolution of a inspiraling BBH.  We
compute the gas-dominated chirp signals in \S\ref{sec:signal} and show that
they resemble the chirp signals from more massive BBHs in vacuum environments.
In \S\ref{sec:MF}, we employ the ``matched-filtering'' technique to quantify
the similarity between the chirp signals in the cases with and without gas.
Finally, in \S\ref{sec:Dis} we discuss the detectability of the fake massive
BBHs in the LISA band.

\section{Chirp signal and the effect of gas}\label{sec:theory}

We focus on the early inspiral phase because the majority of the LISA BBHs are
in this evolutionary phase. Without gas, the semi-major axis $a$
of a BBH decays approximately as
\begin{equation}
	\dot{a}=\dot{a}_{\rm gw}:=\frac{64}{5}\frac{G^3m_1m_2m_{12}}{c^5a^3}\label{eq:adotgw}
\end{equation}
\citep[][assuming near-Keplerian circular orbits]{peters64}, where the dot
symbol denotes the time derivative, $G$ is the gravitational constant, $c$ is
the speed of light, $m_1$ and $m_2$ are the masses of the two BH members, and
$m_{12}=m_1+m_2$ is the total mass of the binary.  During the orbital decay,
the GW frequency, which can be calculated with $f=\pi^{-1}(Gm_{12}/a^3)^{1/2}$,
increases at a rate of $\dot{f}\propto f^{11/3}$.  Such a characteristic
signature is called the ``chirp signal''.  From it, one can derive a
characteristic mass scale
\begin{equation}
{\cal
M}=\frac{c^3}{G}\,\left(\frac{5f^{-11/3}\dot{f}}{96\pi^{8/3}}\right)^{3/5}
=\frac{(m_1m_2)^{3/5}}{(m_1+m_2)^{1/5}},\label{eq:M}
\end{equation}
which is known as the ``chirp mass''. It uniquely determines the time evolution
of $f$. 

From the chirp signal one can also derive the distance $d$ of the BBH
\citep{schutz86}.  This is because from $f$ and $\dot{f}$ one can infer the
energy-loss rate of the orbit, $\dot{E}\propto f^{-4}\dot{f}^2$, which also
equals the GW power.  In addition, the frequency $f$ and the GW amplitude $h$,
together, determine a flux $S\propto h^2f^2$ which is the GW flux.  From the
power and the flux, one can derive the distance of the source
\begin{equation}
d=\frac{4G}{c^2}\frac{{\cal M}}{h}\left(\frac{G}{c^3}\pi\,f\,{\cal M}\right)^{2/3}.\label{eq:d}
\end{equation}

If the BBH is at a cosmological distance, the mass and distance encoded in the
chirp signal will have slightly different meanings.
First, both $f$ and $\dot{f}$ will be
distorted by the redshift so that the observed frequency becomes
$f_o=f(1+z)^{-1}$ and the chirping rate appears to be 
$\dot{f}_o=\dot{f}(1+z)^{-2}$. As a result, the chirp mass that
one will derive from the redshifted GW signal becomes
\begin{equation}
{\cal
M}_o:=\frac{c^3}{G}\,\left(\frac{5f_o^{-11/3}\dot{f_o}}{96\pi^{8/3}}\right)^{3/5}={\cal M}(1+z).\label{eq:M_o}
\end{equation}
This apparent chirp mass is bigger than the intrinsic one by a redshift factor
$1+z$. Second, the GW amplitude will be determined by the transverse comoving
distance $d_C$ in the following way,
\begin{equation}
h_o=\frac{4G}{c^2}\frac{{\cal M}}{d_C}\left(\frac{G}{c^3}\pi\,f\,{\cal M}\right)^{2/3}.\label{eq:h_o}
\end{equation}
If one uses the observed $f_o$, $\dot{f}_o$, and $h_o$ to infer a distance
$d_o$, one will get
\begin{equation}
d_o=\frac{4G}{c^2}\frac{{\cal M}_o}{h_o}\left(\frac{G}{c^3}\pi\,f_o\,{\cal M}_o\right)^{2/3}=d_C(1+z).\label{eq:d_o}
\end{equation}
Such a distance is identical to the luminosity distance $d_L$ in a
$\Lambda$CDM cosmology.

Gas will accelerate the orbital shrinkage and, in this way, affect the chirp
signal. Because of the gas friction \citep[or viscosity, e.g.,][]{haiman09}, a BBH
would shrink at a faster rate of $\dot{a}=\dot{a}_{\rm gw}+\dot{a}_{\rm gas}$,
where the additional term, $\dot{a}_{\rm gas}<0$, is due to gas.
Correspondingly, $\dot{f}_o$ increases more rapidly, as
\begin{equation}
	\frac{\dot{f}_o}{f_o}=-\frac{3}{2}\left[\frac{\dot{a}_{\rm gw}+\dot{a}_{\rm gas}}{a(1+z)}\right].\label{eq:fdottot}
\end{equation}

The apparent increase of $\dot{f}_o$ will lead to an overestimation of the
mass of the BBH, as well as the distance. To see this effect, it is useful to
first define an acceleration factor
\begin{equation}
\Gamma:=\dot{a}_{\rm gas}/\dot{a}_{\rm gw}.
\end{equation}
From this definition, it follows that
$\dot{f}_o=(1+\Gamma)(1+z)^{-2}\dot{f}$, i.e., the chirp signal evolves faster by
a factor of $1+\Gamma$.  Finally, by revisiting Equations~(\ref{eq:M_o}) and
(\ref{eq:d_o}), we find that the apparent mass and distance become
\begin{align}
	{\cal M}_o&=(1+\Gamma)^{3/5}{\cal M}(1+z),\label{eq:calM_o}\\
	d_o&=(1+\Gamma)\,d_C(1+z)=(1+\Gamma)\,d_L.\label{eq:bigdo}
\end{align}

The factor $\Gamma$ in general is a function of $a$, because both $\dot{a}_{\rm
gw}$ and $\dot{a}_{\rm gas}$ depend on $a$. Such a dependence has two
consequences.  (i) If the semi-major axis changes substantially during the
observational period, one would see a significant variation of ${\cal M}_o$ and
$d_o$ with time. This result is inconsistent with the dynamical evolution of a
BBH in a vacuum, and hence can be used to prove the presence of a environmental
factor, such as gas.  (ii) Otherwise, if $a$ evolves very slowly, i.e., the
corresponding evolutionary timescale $a/|\dot{a}_{\rm gw}+\dot{a}_{\rm gas}|$
is much longer than the observational period $T_{\rm obs}$, the acceleration
factor $\Gamma$ would be more or less a constant. In this case, the measurement
of ${\cal M}_o$ and $d_o$ would be relatively consistent during the
observational period, and both values would be greater than the intrinsic ones.  

For LIGO/Virgo, the relevant BBHs are normally in the first case because the
signal is typically less than a second but each LIGO/Virgo observing run lasts
several weeks to several months.  This is why the previous studies found that
the gas effect could be discerned in the LIGO/Virgo waveforms (see \S\ref{sec:intro}).  For LISA, however, the majority of the in-band BBHs belong
to the latter case because a BBH could dwell in the band for as long as
millions of years but the canonical mission duration of LISA is only $4-5$
years.  During such a short observing time, $\Gamma$ is almost constant so that
discerning the gas effect is more difficult. An overestimation of the mass and
distance becomes more likely.

\section{Hydrodynamics versus GW radiation}\label{sec:compare}

To evaluate the efficiency of the hydrodynamical drag,
we compare the GW radiation timescale, defined as $\tau_{\rm
gw}:=|a/\dot{a}_{\rm gw}|$, and the hydrodynamical timescale, defined as
$\tau_{\rm gas}:=|a/\dot{a}_{\rm gas}|$. Following these definitions,
the acceleration factor $\Gamma$ equals $\tau_{\rm gw}/\tau_{\rm gas}$.
Because the most sensitive band of LISA is around $3$ mHz, the corresponding
BBHs have a typical semi-major axis of
\begin{equation}
a=\left(\frac{Gm_{12}}{\pi^2 f^2}\right)^{1/3}\simeq0.0021
\left(\frac{m_{12}}{20~M_\odot}\right)^{1/3}
\left(\frac{f}{3\,{\rm mHz}}\right)^{-2/3}\,{\rm AU}.
\end{equation}
According to Equation~(\ref{eq:adotgw}), without gas these binaries have a typical 
evolutionary timescale of 
\begin{align}
	\tau_{\rm gw}&=\frac{5}{64}\frac{c^5a^4}{G^3m_1m_2m_{12}}\\
&\simeq\frac{9.1\times10^3}{q(1+q)^{-1/3}}
\left(\frac{m_1}{10\,M_\odot}\right)^{-5/3}
\left(\frac{f}{3\,{\rm mHz}}\right)^{-8/3}\,{\rm years},
\end{align}
where $q$ denotes the mass ratio $m_2/m_1$ of the two BHs (we assume $m_1\ge m_2$).

As for $\tau_{\rm gas}$, we first use the hydrodynamic drag derived in 
\citet{ostriker99,sanchez99} to estimate its value. 
The drag force on the secondary BH ($m_2$) is calculated with
$F\sim4\pi\rho(Gm_2/v)^2$, where $\rho$ is the mass density of the background gas
and $v$ is the Kepler velocity of the secondary. 
For circular orbits, we have $v\propto a^{-1/2}$, so that 
$a/\dot{a}_{\rm gas}=-v/(2\dot{v})$. Moreover, since $|\dot{v}|=F/m_2$, we   
derive that
\begin{equation}
	\tau_{\rm gas}=\frac{m_2v}{2F}\simeq\frac{1.1\times10^4}{q(1+q)^{2}}
\left(\frac{n}{10^{16}\,{\rm cm^{-3}}}\right)^{-1}
	\left(\frac{f}{3\,{\rm mHz}}\right) \,{\rm yrs},
\end{equation}
where $n$ is the number density of hydrogen atoms in the gas background.
It follows that
\begin{align}
\Gamma\simeq4.3\,\left(\frac{1+q}{2}\right)^{7/3}
\left(\frac{n}{10^{16}\,{\rm cm^{-3}}}\right)
\left(\frac{m_1}{10\,M_\odot}\right)^{-5/3}
	\left(\frac{f}{3\,{\rm mHz}}\right)^{-11/3}.\label{eq:Gamm}
\end{align}

\begin{figure}
\plotone{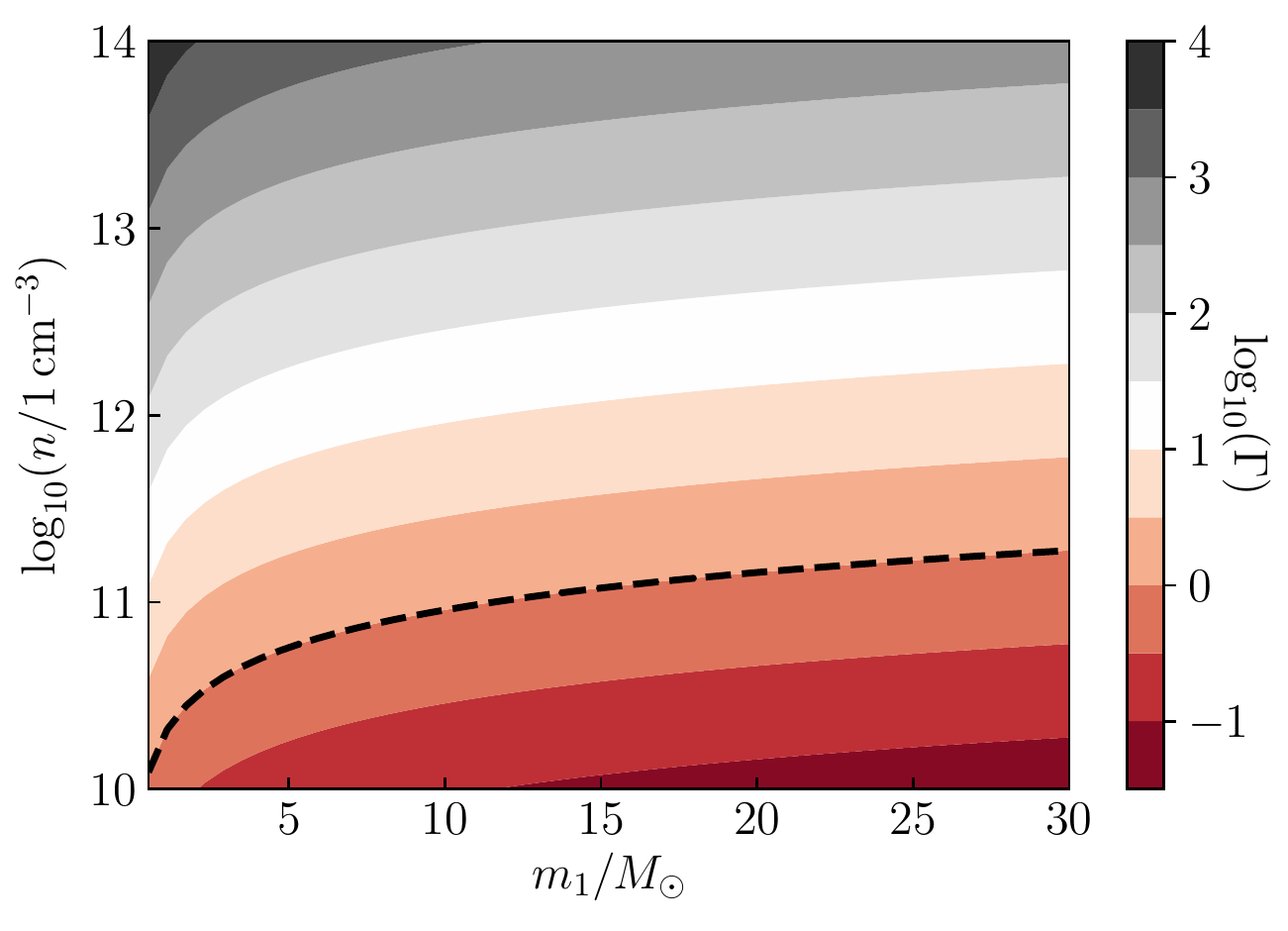}
	\caption{Dependence of the acceleration factor, calculated in 
	Eq.~(\ref{eq:Gamm2}), on the mass of the primary BH, $m_1$, and the
	gas density $n$. The other model parameters are set to
	$f=3$ mHz, $q=1$, and $c_s=10^2\,{\rm km\,s^{-1}}$. The black dashed curve
	marks the location where ${\Gamma}=1$.} \label{fig:Gamm}
\end{figure}

In the above derivation of $\tau_{\rm gas}$, it is assumed that the gas background is
homogeneous and the small body is moving in a straight line. However, for the
BHs in binaries, which move along Keplerian orbits, it has been shown that the
formula for the drag force will be modified, because the shape of the density
wake is different \citep{sanchez01,escala04,kim07,kim08}.  More recently,
\citet{antoni19} showed that when $a$ is smaller than the Bondi 
accretion radius $R_{\rm
acc}=Gm_{12}/c_s^2$ ($c_s$ being the sound speed of the gas medium), 
the gas density close to the binary will be much
higher than the background density due to accretion.
This is normally the case for those BBHs embedded in common envelopes
and AGN accretion disks.
If we use $\tilde{n}\sim
n\,(R_{\rm acc}/a)^{3/2}$ to correct the gas density around the binary \citep{bondi52,antoni19}, we find that the timescale due to hydrodynamical drag becomes 
\begin{align}
	{T}_{\rm gas}\simeq&8.5\times10^{4}q^{-1}(1+q)^{-3}
\left(\frac{n}{10^{11}\,{\rm cm^{-3}}}\right)^{-1}\nonumber\\
&\times
\left(\frac{m_1}{10\,M_\odot}\right)^{-1}
	\left(\frac{c_s}{10^2\,{\rm km\,s^{-1}}}\right)^{3}\,{\rm years}.\label{eq:Tgas}
\end{align}
Note that the new timescale does not depend on $a$ or $f$.  A similar result
can be found in \citep[][see their Fig. 2]{bartos17}.  Moreover, in the last
equation we have rescaled $n$ with $10^{11}\,{\rm cm^{-3}}$.  Given this
timescale, we derive that
\begin{align}
	{\Gamma}&\simeq1.1\,\left(\frac{1+q}{2}\right)^{10/3}
\left(\frac{n}{10^{11}\,{\rm cm^{-3}}}\right)\nonumber\\
&\times\left(\frac{m_1}{10\,M_\odot}\right)^{-2/3}
\left(\frac{c_s}{10^2\,{\rm km\,s^{-1}}}\right)^{-3}
	\left(\frac{f}{3\,{\rm mHz}}\right)^{-8/3}.\label{eq:Gamm2}
\end{align}
For illustrative purposes, we show in Figure~\ref{fig:Gamm} the dependence of
the ${\Gamma}$ computed in the last equation on $m_1$ and $n$. The black dashed
curve marks the location where ${\Gamma}=1$.  Above it, gas dominates the
dynamical evolution of a BBH, and hence the chirp signal is determined by gas
dynamics, not GW radiation. 

Equations~(\ref{eq:Gamm}) and (\ref{eq:Gamm2}) suggest that for LIGO/Virgo
BBHs, which typically have $m_1\sim10\,M_\odot$ and $f\sim10^2$ Hz, the gas
effect is negligible unless the gas density $n$ is orders of magnitude higher
than $10^{16}\,{\rm cm^{-3}}$. This is the reason that for LIGO/Virgo sources,
significant gas effect is expected only in stellar cores, where $n$ could be as
high as $(10^{28}-10^{31})\,{\rm cm^{-3}}$ \citep[e.g.][]{fedrow17}.  For LISA
sources with $f\sim3$ mHz, however, gas effect is already important when
$n\sim10^{16}\,{\rm cm^{-3}}$ according to Equation~(\ref{eq:Gamm}) or
$n\sim10^{12}\,{\rm cm^{-3}}$ according to Equation~(\ref{eq:Gamm2}).  These
two characteristic densities can be found, respectively, in common envelopes
and AGN accretion disks \citep[e.g.][]{antoni19}.  Therefore, the gaseous
environments common for BBHs would affect the LISA signals more than the
LIGO/Virgo ones.

\section{Faking a chirp signal}\label{sec:signal}

Having understood the effect and the relative importance of gas, we now compute
the chirp signal of a BBH embedded in a gaseous environment.  In the following,
we assume $z=0$ for simplicity ($f_o=f$). When there is no gas, we calculate
the time derivative of the GW frequency ($\dot{f}_{\rm gw}$) using a 3.5
post-Newtonian (PN) approximation presented in \citet{sathyaprakash09}.  When
gas is present, we have $\dot{f}_o=\dot{f}_{\rm gw}+\dot{f}_{\rm gas}$, where
$\dot{f}_{\rm gas}$ is calculated with $\dot{f}_{\rm gas}=-3f/(2\tau_{\rm
gas})$ according to Equation~(\ref{eq:fdottot}).  In this way, the gas effect
is included in the model through a parameter $\tau_{\rm gas}$. 

\begin{figure}
\begin{center}
\includegraphics[clip,width=0.5\textwidth]{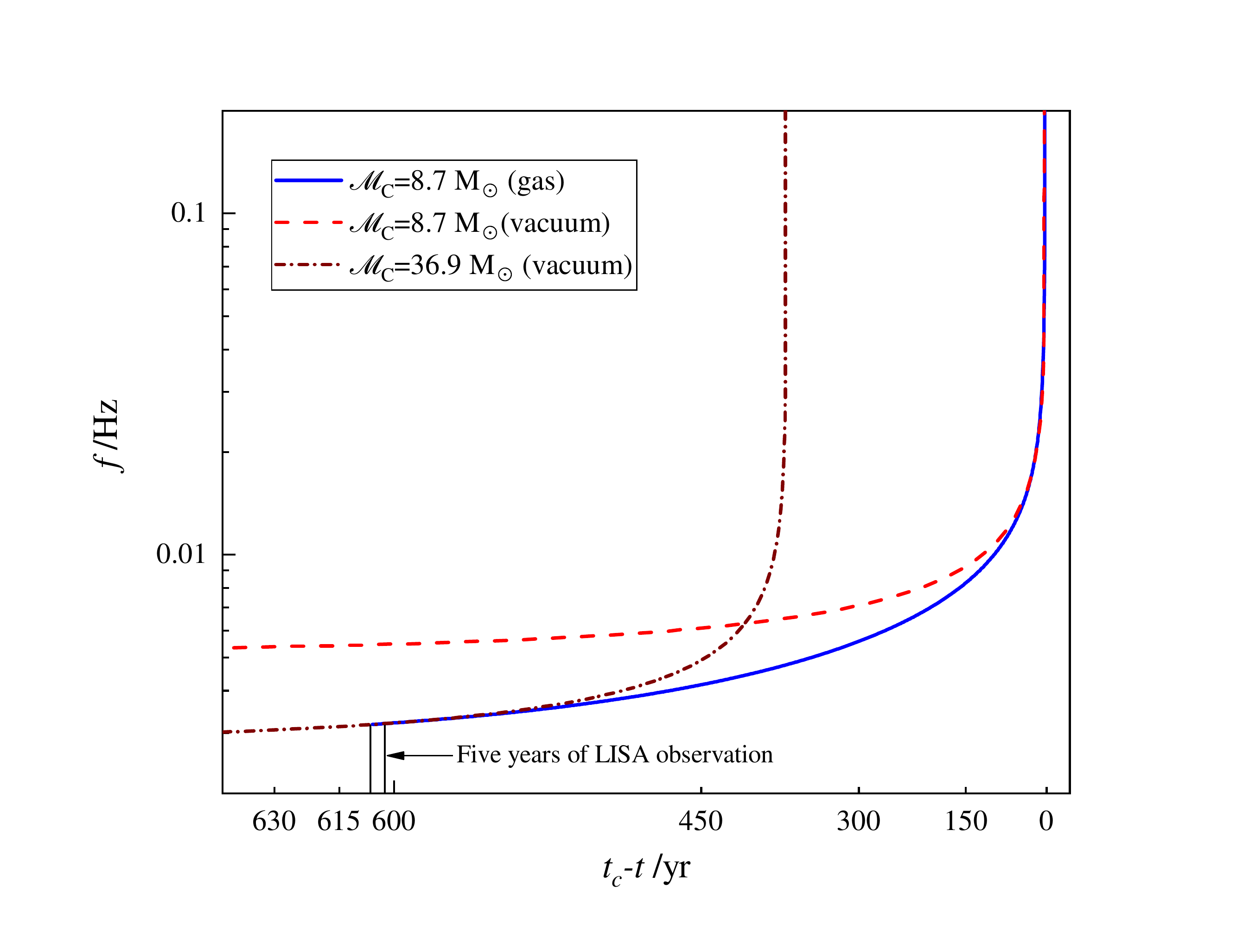}
\caption{Evolution of the GW frequency as a function of the time before coalescence.
	The blue solid curve
	shows a $10M_\odot-10M_\odot$ circular BBH embedded in
	a gaseous environment. The model parameters are
	chosen such that ${\Gamma}=10$ when the GW frequency is $f=3$
	mHz (see Eq.(\ref{eq:Gamm2})). The two vertical lines mark the typical LISA observational window
	of five years. The red dashed curve shows the chirp signal of the same
	BBH but placed in a vacuum. The coalescence times of the previous
	two BBHs are aligned
	for easier comparison.  The dot-dashed curve shows another BBH
merging in vacuum but with higher masses. It is offset in time so that the GW
frequency in the LISA observational window is also $3$ mHz.  } \label{fig:long}
\end{center}
\end{figure}

Figure~\ref{fig:long} compares the long-term evolution of the chirp signal in
the cases with and without the gas drag.  The blue solid curve corresponds to a
$10M_\odot-10M_\odot$ BBH (${\cal M}\simeq8.7\,M_\odot$) embedded in a gaseous
medium. The two BHs coalesce at the time $t=t_c$.  The hydrodynamical
timescale $\tau_{\rm gas}$ is computed according to Equation~(\ref{eq:Tgas}) and
the model parameters are chosen such that $\tau_{\rm gw}/\tau_{\rm gas}=10$ when
$f=3$ mHz.  This chirp signal around $f=3$ mHz, according to
Equation~(\ref{eq:calM_o}), should resemble a more massive binary with a chirp
mass of $(1+\Gamma)^{3/5}{\cal M}\simeq37\,M_\odot$ in a vacuum.  The chirp
signal of the latter more massive BBH is shown as the dot-dashed curve, and we
can see that, indeed, at $f=3$ mHz it matches the signal of the smaller binary
embedded in gas.  Eventually, the two signals diverge, since $\Gamma$ is
decreasing, but the divergence appears more than one hundred years later.  If
we focus on the LISA observational window of five years (marked by the two
vertical lines), the two signals are almost identical.  Finally, during the
coalescence, the blue curve recovers the chirp signal of a
$10M_\odot-10M_\odot$ binary in a vacuum (red dashed curve), because GW
radiation predominates during the merger. 

\begin{figure}
\begin{center}
\includegraphics[clip,width=0.5\textwidth]{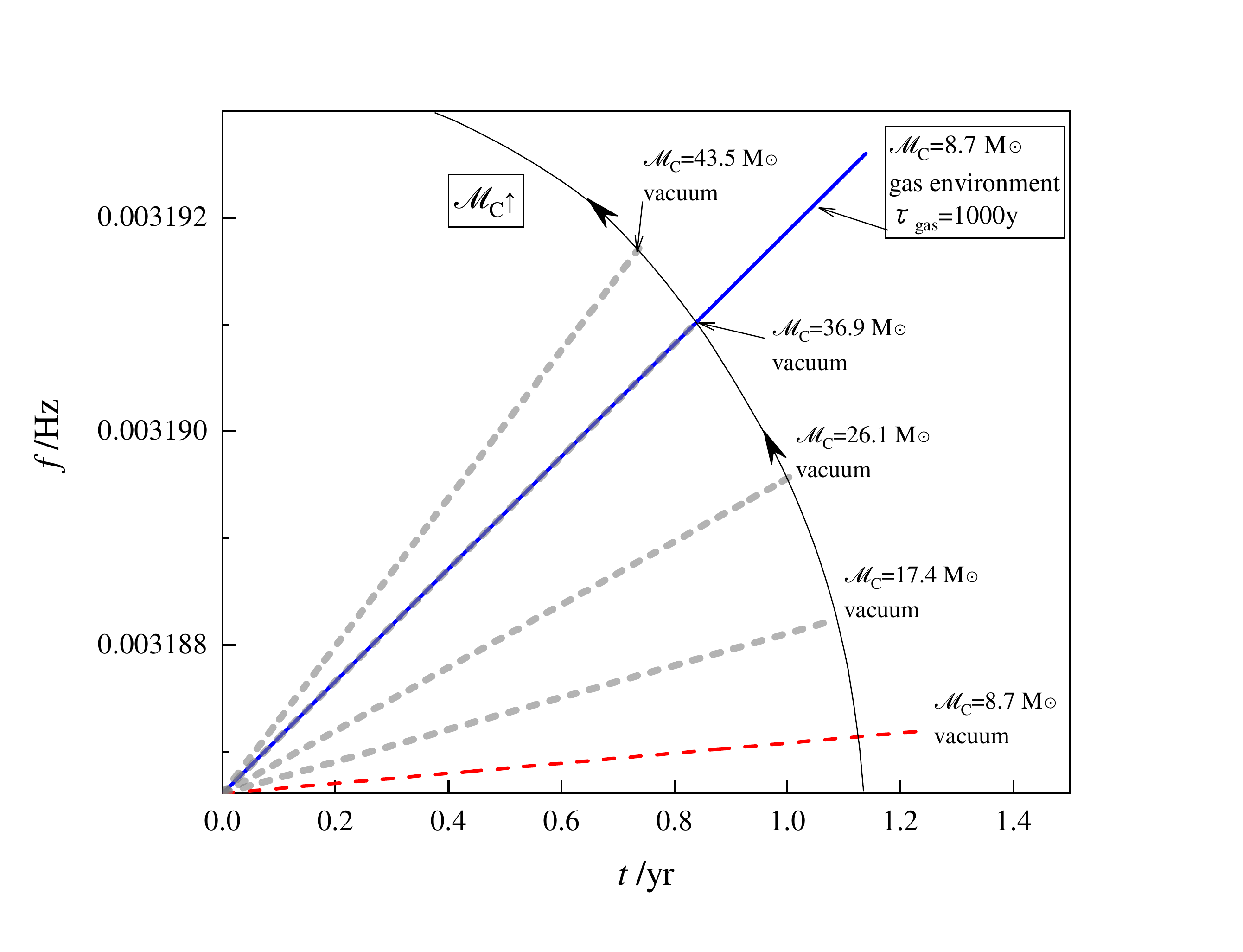}
\caption{Chirp signals in the LISA observational window. The blue solid curve 
	is computed with the gas effect and the
	red dashed one without. The model parameters are the same as in Fig.~\ref{fig:long}. The gray dashed curves show the dependence of the chirp signal
	on the chirp mass when there is no gas.} \label{fig:short}
\end{center}
\end{figure}

To see more clearly the chirp signal in the observation window of LISA, we show
in Figure~\ref{fig:short} the evolution of $f$ during a period of $1-2$ years,
around the moment when $f$ is approximately $3$ mHz. Now the chirp signals look
like straight lines because the observational period is orders of magnitude
shorter than the evolutionary timescales of the BBHs.  Although the variation of
$f$ is small during the observation period, it is detectable by LISA because
LISA's resolution is approximately $10^{-8}(1\,{\rm yr}/T_{\rm obs})(10/ {\rm
SNR})$ Hz \citep{seto02}.  Comparing the blue solid and the red dashed lines,
we see that the presence of gas increases the slop the chirp signal. The
steeper line resembles the chirp signal of a more massive BBH in a vacuum with
a chirp mass of $37\,M_\odot$.

We have seen that with or without gas, the chirp signals in the LISA band are
almost straight lines and are relatively featureless compared to those in the
LIGO/Virgo band.  This is the main reason that in this band a BBH in a gaseous
environment could be misidentified as a more massive binary in a vacuum.

\section{Misidentification}\label{sec:MF}

LISA uses a technique called ``matched filtering'' to search for BBHs in the
data stream \citep{Finn92,cutler94}. In this section, we will show that this
method cannot distinguish light BBHs embedded in certain gaseous environments
from those massive ones in vacuum.

\subsection{Matched filtering}

Given two waveforms, $h_1$ and $h_2$, their similarity is quantified by a
``fitting factor'' (FF), which is defined as
\begin{equation}
{\rm FF}=\frac{\left<h_1|h_2\right>}{\sqrt{\left<h_1|h_1\right>\left<h_2|h_2\right>}}.
\end{equation}
The term $\left<h_1|h_2\right>$ means an inner product of
\begin{equation}
\left<h_1|h_2\right>=2\int_0^\infty\frac{\tilde{h}_1(f)\tilde{h}_2^*(f)+\tilde{h}_1^*(f)\tilde{h}_2}{S_n(f)}df,
\end{equation}
where the tilde symbols stand for the Fourier transformation, the stars stand
for the complex conjugation, and $S_n(f)$ is the spectral noise density of LISA
\citep{klein16}. Identical waveforms have ${\rm FF}=1$.

In our problem, $h_1(t)$ is the chirp signal of a BBH embedded in a gaseous
environment, and $h_2(t)$ is the waveform of a inspiraling BBH in a vacuum.  By
tuning the parameters of $h_2$, we want to maximize the FF. We follow
\citet{cutler94} and compute the waveforms using
\begin{equation}
	h(t)=\frac{Q(\theta, \varphi, \psi, \iota) \mu M}{d_L\, a(t)} \cos \left(\int 2 \pi f d t\right),
\end{equation}
where $Q(\theta, \varphi, \psi, \iota)$ is a function depending on the sky
location and orientation of the BBH. In the integrand, the frequency $f$ is a
function of $a$. It is computed using the 3.5 PN approximation for $h_2$
\citep{sathyaprakash09} and using the gas model described in
\S\ref{sec:signal} for $h_1$.

Because the two evolutionary timescales $\tau_{\rm gas}$ and $\tau_{\rm gw}$ are both
much longer than the observational period of LISA, $f$ is almost a constant in
our model. In this case the computation of the inner product
$\left<h_1|h_2\right>$ can be performed in the time domain and the calculation of the
FF can be simplified.  First, the noise curve $S_n(f)$ can be taken out of the
integration because of the small variation of $f$, so that
\begin{equation}
	\left<h_{1} | h_{2}\right>\approx \frac{2}{S_n} \int_{0}^{\infty} [\tilde{h}_{1}^{*}(f) \tilde{h}_{2}(f)+\tilde{h}_{1}(f) \tilde{h}_{2}^{*}(f)] d f.
\end{equation}
Second, using Parseval's theorem, we further derive
\begin{equation}
	\left <h_{1} | h_{2}\right >\approx \frac{4}{S}\int_{0}^{\infty} h_{1}(t) h_{2}(t)d t.
\end{equation}
Finally, the FF can be written as
\begin{equation}
	{\rm FF}=\frac{\int_{0}^{\infty} h_{1}(t) h_{2}(t)d t}{\sqrt{\int_{0}^{\infty} h_{1}(t) h_{1}(t)d t\int_{0}^{\infty} h_{2}(t) h_{2}(t)d t}}.
\end{equation}
Note that there is no more dependence on $d_L$, $Q(\theta, \varphi, \psi,
\iota)$, or $S_n(f)$, because they all cancel out. Given $h_1$, i.e., the
signal, we want to find a template $h_2$ that maximizes the FF. The parameter
space in which we conduct this search is $({\cal M}_c,\,q,\,\phi)$,
where $\phi$ is the initial phase.

\subsection{Examples}\label{sec:examp}

\begin{figure}
\begin{center}
\includegraphics[clip,width=0.5\textwidth]{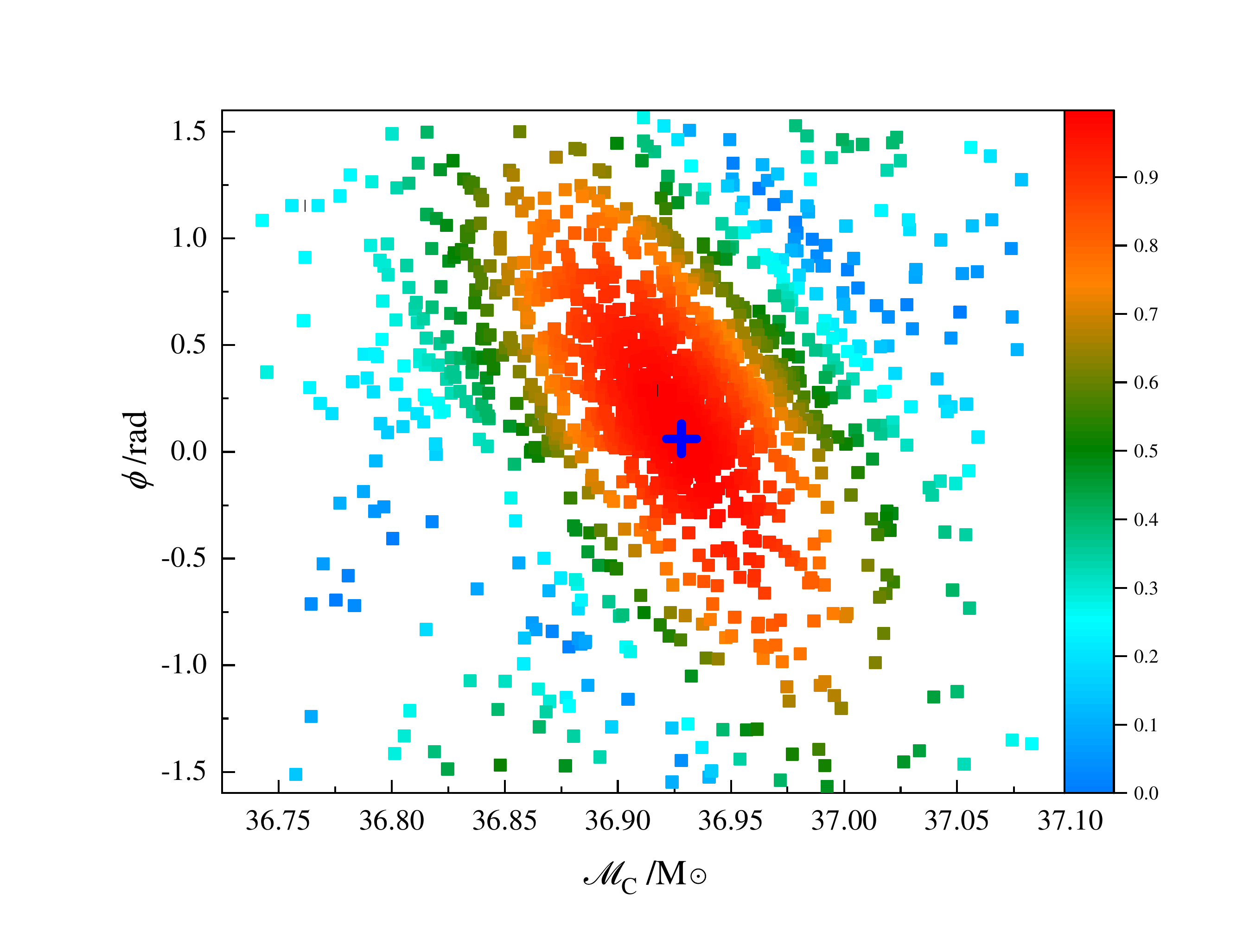} \caption{Dependence of
	the FF on the chirp mass ${\cal M}_c$ and initial phase $\phi$ of the
	template $h_2$. The dependence on $q$ is relatively weak and
	is not shown here.  The signal $h_1$ is generated using the
	parameters ${\cal M}_c=8.7\,M_\odot$, $q=0.7$, and $\tau_{\rm gas}=10^3$
	years. The plus symbol marks
	the location of the maximum FF, which is offset from the
	real chirp mass ($8.7\,M_\odot$).}\label{fig:heatmap} 
\end{center} 
\end{figure}

Figure~\ref{fig:heatmap} shows one example of our search. The signal
is generated using a BBH with ${\cal M}_c=8.7\,M_\odot$, $q=0.7$, and a
hydrodynamical timescale of  $\tau_{\rm gas}=10^3$ years.  Initially, the GW
frequency is $f=3$ mHz, and the corresponding GW radiation timescale is about
$10^4$ years.  In this particular example the observational period is set to
$T_{\rm obs}=1.25$ years, but later we will show the FF for different $T_{\rm
obs}$.  We match the signal using the templates developed for vacuum BBHs,
i.e., the 3.5 PN approximation described above.  We use a simulated annealing
algorithm to search for the highest FF in the parameter space of $({\cal
M}_c,\,q,\,\phi)$.  The best FF is found at a chirp mass of ${\cal
M}_c\simeq37\,M_\odot$. It is offset from the real chirp masses by a factor of
about $4.2$, which is consistent with our Equation~(\ref{eq:calM_o}). This
result confirms our prediction that ignoring the gas effect could result in a
significant overestimation of the mass of a LISA BBH.

\begin{figure}
\begin{center}
\includegraphics[clip,width=0.5\textwidth]{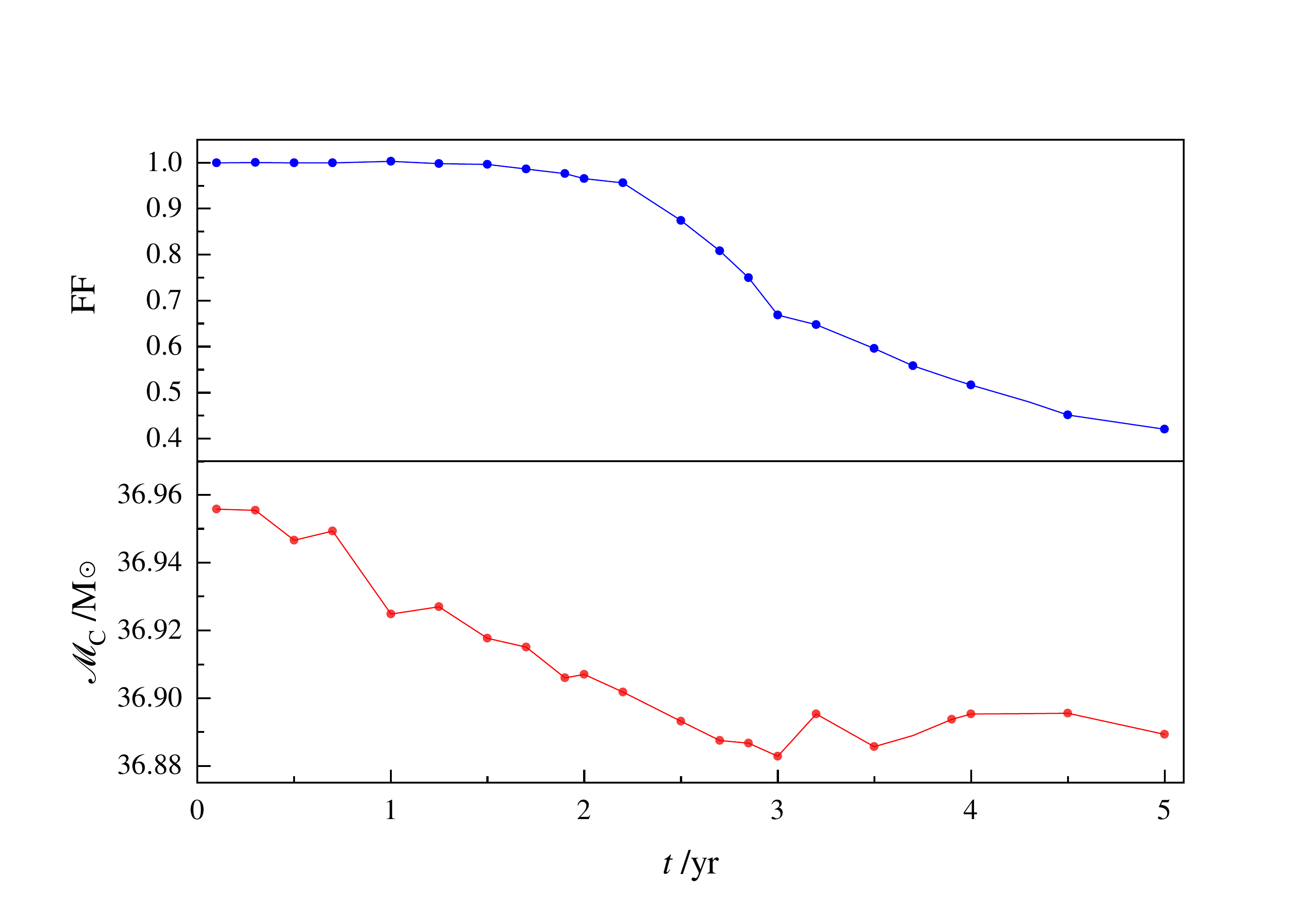}
\caption{Variation of the maximum FF and the best-fit chirp mass with 
	the observational period. The real chirp mass of the BBH is $8.7\,M_\odot$.} \label{fig:FFsingle}
\end{center}
\end{figure}

We note that the best FF and the corresponding best-match ${\cal M}_c$ are both
functions of $T_{\rm obs}$. Figure~\ref{fig:FFsingle} shows such a dependence
on time. The model parameters are the same as in Figure~\ref{fig:heatmap}.
During the first $1-2$ years, the FF remains close to $1$ and afterwards
decreases with time. The FF deteriorates on a long timescale because the
perturbation on the GW phase by the gas effect is accumulative.  For the
best-fit ${\cal M}_c$, it decreases with time but remains close to
$37\,M_\odot$.

Whether or not LISA is able to distinguish $h_1$ from $h_2$ depends on not only 
the FF but also the SNR, since noise also plays a role. The SNR is defined as 
\begin{equation}\label{eq:SNR2}
{\rm SNR}^2:=\left<h|h\right>.
\end{equation}
According to \citet{lindblom08}, two waveforms are distinguishable when 
$\left<\delta h|\delta h\right> >1$, where $\delta h:=\tilde{h}_1(f)-\tilde{h}_2(f)$.
This criterion can be simplified in our problem because we are often in a
situation where $h_1\simeq h_2$. In this case we have
${\rm SNR}^2\simeq\left<h_1|h_1\right>\simeq\left<h_2|h_2\right>$, and the criterion
reduces to
\begin{equation}
{\rm
	FF}<1-1/(2\,{\rm SNR}^2).\label{eq:FF}
\end{equation}
If we take ${\rm SNR}\simeq10$ as the threshold for LISA to claim a detection,
the corresponding criterion of distinguishing two different waveforms becomes
${\rm FF}<0.995$.

According to this criterion, the gas and vacuum waveforms used in
Figure~\ref{fig:FFsingle} are indistinguishable during the first $1-2$ years of
observation. Only in the third year could one start to tell the difference and
prove that the signal is not produced by a massive BBH of ${\cal
M}_c\simeq37\,M_\odot$ residing in a vacuum environment.

\begin{figure}
\begin{center}
\includegraphics[clip,width=0.5\textwidth]{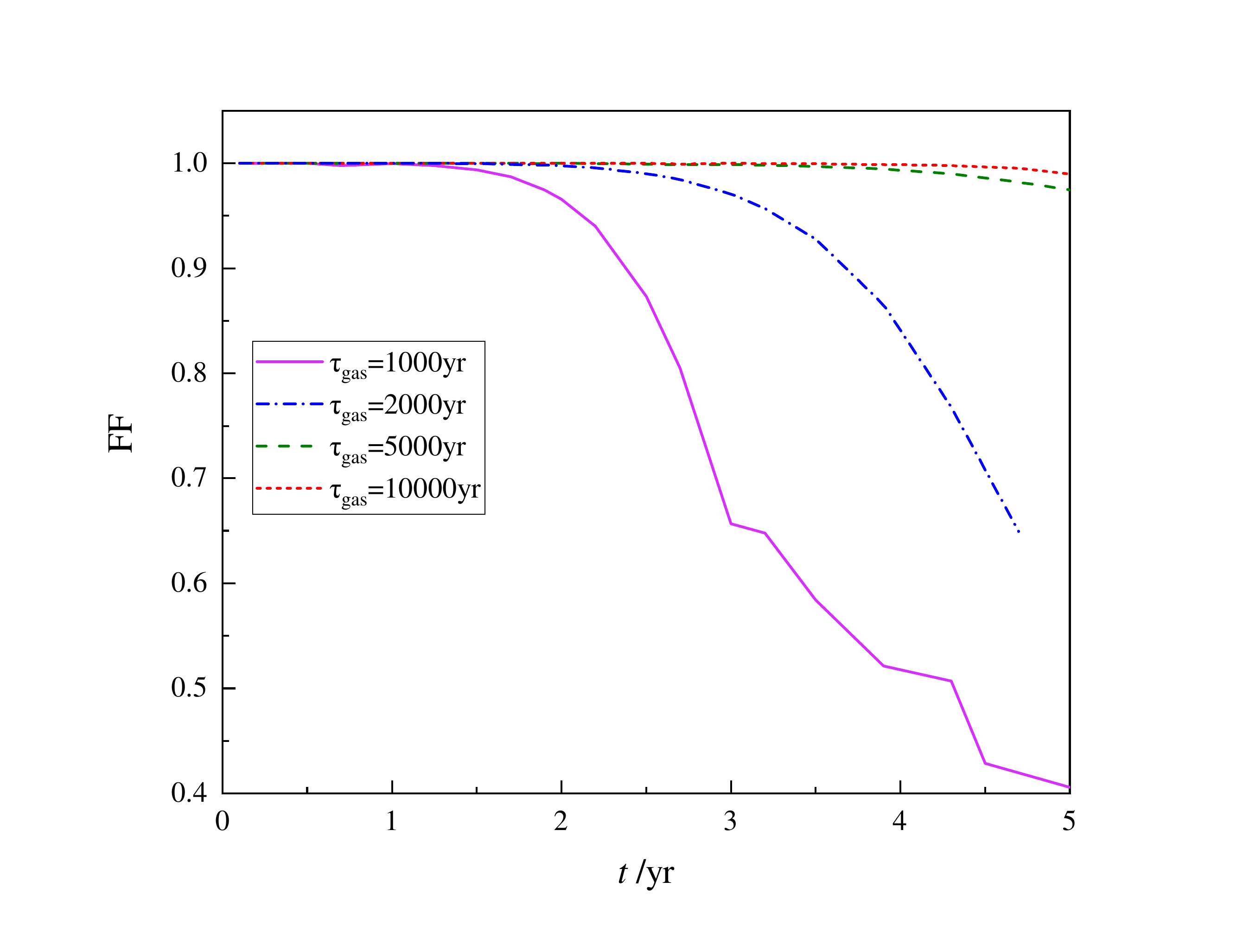}
	\caption{Variation of the maxim FF with time assuming different values of 
	$\tau_{\rm gas}$.} \label{fig:FFmultiple}
\end{center}
\end{figure}

For completeness, we show in Figure~\ref{fig:FFmultiple} the FF derived
assuming different values for $\tau_{\rm gas}$.  The other model parameters are
the same as in Figure~\ref{fig:heatmap}.  As $\tau_{\rm gas}$ increases, the FF
gets better at later times because the gas effect becomes weaker. We note that
when $\tau_{\rm gas}\ga5,000$ years, the FF is better than $0.995$ for almost
five years, which is equivalent to the canonical mission duration of LISA.  As
a result, LISA may identify our BBHs of a chirp mass of $8.7\,M_\odot$ as more
massive binaries.  The measured chirp mass is approximately $18\,M_\odot$ when
$\tau_{\rm gas}=5,000$ years and $14\,M_\odot$ when $\tau_{\rm gas}=10^4$
years. In both cases, the overestimation of the mass, and hence the distance
(see Eq.~(\ref{eq:bigdo})), is substantial.

\section{Discussions}\label{sec:Dis}

We have seen that those BBHs with $\Gamma=\tau_{\rm gw}/\tau_{\rm gas}\ga1$
could be misidentified by LISA as more massive binaries residing in vacuum
environments. To what distance could LISA detect such fake massive binaries?  The
standard way of addressing this question is to derive the maximum distance at
which the SNR drops to a threshold, say ${\rm SNR}=10$. This distance is known
as the ``detection horizon''. In the case without gas, the detection horizon
has been derived in several works assuming different LISA configurations. For
example, \citet{kyutoku17} showed that 
\begin{align}
	d_L\simeq&13\,\left(\frac{{\cal M}_c}{10\,M_\odot}\right)^{5/3}
	\left(\frac{T_{\rm obs}}{5\,{\rm yr}}\right)^{1/2}
	\left(\frac{{\rm SNR}}{10}\right)^{-1}\nonumber \\
&\times
\left(\frac{S_n(f)}{10^{-40}\,{\rm Hz^{-1}}}\right)^{-1/2}
	\left(\frac{f}{3\,{\rm mHz}}\right)^{2/3}\,{\rm Mpc}\label{eq:SNR}
\end{align}
for the N2A5 configuration of LISA.  Here ${\cal M}_c$ and $d_L$ refer
to the real chirp mass and real distance of the source.  

In fact, the last equation is valid also in the case with gas. This is so
because of the following reasons. (i) The stationary phase approximation
\citep{thorne87} in which the last equation is derived remains valid, since the
evolutionary timescale $f/\dot{f}$ for the frequency is much longer than the GW
period $1/f$. We note that $\dot{f}$ here stands for the observed frequency. 
We dropped the subscript $o$ in $\dot{f}_o$
for simplicity. (ii) In this approximation, the characteristic amplitude defined
as $h_c(f):=2f\tilde{h}(f)$ becomes proportional to $A\dot{f}^{-1/2}$, where
$A$ is a function of ${\cal M}_c$, $q$, $f$, and $d_L$, as well as the
sky location and orientation of the binary. Noticing that $\dot{f}=\dot{f}_{\rm gw}+\dot{f}_{\rm gas}>\dot{f}_{\rm gw}$, we find that gas in general reduces the
characteristic amplitude at any frequency. The reduction is due to a
faster drift of the signal in the frequency domain.
(iii) Using $h_c$,  we can
rewrite the SNR defined in Equation~(\ref{eq:SNR2}) as 
\begin{equation}
	{\rm SNR}^2=\int_{f_1}^{f_2}\frac{|h_c(f)|^2}{f^2S_n(f)}df,
\end{equation}
where $f_1$ and $f_2$ denote the minimum and maximum frequencies during the
observational period. Because $\Delta f=f_2-f_1\ll f$, the integration becomes
proportional to $|h_c|^2\Delta f/(f^2S_n)$. (iv) In our problem we have $T_{\rm
obs}\ll |f/\dot{f}|$. Therefore, we can write $\Delta f\simeq \dot{f}T_{\rm
obs}$. Again, by noticing that $\dot{f}>\dot{f}_{\rm gw}$,
we find that $\Delta f$ is broader when gas is present.
(v) Finally, the $\dot{f}^{-1}$ from the term $|h_c|^2$ cancels the $\dot{f}$
from the term $\Delta f$, so that the SNR does not depend on $\dot{f}$.
Physically, this means the reduction of the characteristic amplitude is
compensated by the larger frequency drift. 

The above conclusion that gas does not affect the SNR is derived in the
scenario of LISA observations. It does not apply to LIGO/Virgo because in
the latter case the assumption $T_{\rm obs}\ll |f/\dot{f}|$ is invalid.  In
fact, gas will reduce the SNR for LIGO/Virgo sources by suppressing $|h_c|$.
Nevertheless, the corresponding change of SNR is small because, as has been
explained in \S\ref{sec:compare}, the acceleration factor $\Gamma$ is small
when a BBH enters the LIGO/Virgo band.

Therefore, we can use Equation~(\ref{eq:SNR}) to estimate the detection horizon
and, based on it, discuss the detectability of the fake massive BBHs embedded
in gaseous environment. For ${\cal M}_c=10\,M_\odot$,
the detection horizon corresponding
to a SNR of $10$ is approximately $13$ Mpc, assuming $5$ years of observation.
If ${\cal M}_c=30\,M_\odot$, as the LIGO/Virgo observations tend to suggest
\citep{ligo18a}, the detection horizon elongates to about $80$ Mpc.

To estimate the number of fake massive BBHs within the detection horizon, we
start from the event rate in the LIGO/Virgo band, which is estimated to be
${\cal O}(10^2)\,{\rm Gpc^{-3}\,yr^{-1}}$ \citep{ligo18a}.  According to this
rate, the number of BBHs in the last year before their coalescence is about
$N(\tau=1\,{\rm yr})=100$ per ${\rm Gpc}^{3}$, where $\tau=|a/\dot{a}|$ denotes
the orbital evolutionary timescale.  For the other BBHs at an earlier
evolutionary stage, we can follow the continuity equation \citep[e.g., see Sec.
V in][]{amaro-seoane19} and derive that $dN/d\ln a\propto \tau$. In our
problem, $\tau=(1/\tau_{\rm gw}+\tau_{\rm gas})^{-1}$. At a frequency of $f=3$
mHz, where LISA is the most sensitive, $\tau_{\rm gw}\simeq(1500-9000)$ years
when ${\cal M}_c$ varies from $30\,M_\odot$ to $10\,M_\odot$. (i) Without gas,
$\tau=\tau_{\rm gw}$, and we find that the number density of BBHs at $f\sim3$
mHz is about $(1.5-9)\times10^{5}\,{\rm Gpc}^{-3}$. The number of BBHs inside
the detection horizon is $8-320$. (ii) With gas, the number would be smaller
because $\tau$ is shortened by gas friction. In the extreme case that all BBHs
are embedded in gas, if we assume $\tau_{\rm gas}=10^3$ years, we find that
$\tau\simeq(600-900)$ years when ${\cal M}_c$ varies from $30\,M_\odot$ to
$10\,M_\odot$. Correspondingly, the number density of BBHs at $f\sim3$ mHz is
$(6-9)\times10^{4}\,{\rm Gpc}^{-3}$. The number of BBHs inside the
detection horizon decreases to $0.8-130$, but is not zero. 

Since BBHs embedded in gaseous environments could be common, the effect of
hydrodynamics should be considered more carefully in the waveform modeling.
Otherwise, as our results suggest, LISA may provide a biased demography of
BBHs.  Such a bias may also affect future cosmology studies, given the
possibility of using BBHs as standard sirens to measure cosmological
parameters.

{\it Acknowledgement.}--This work is supported by the NSFC grants No.  11873022
and 11991053. XC is supported partly by the Strategic Priority Research Program
“Multi-wavelength gravitational wave universe” of the Chinese Academy of
Sciences (No. XDB23040100 and XDB23010200). The computation in this work was
performed on the High Performance Computing Platform of the Center for Life
Science, Peking University. 

\bibliographystyle{astroads.bst}

\end{document}